\documentclass[aps,amsmath,amssymb,superscriptaddress,longbibliography]{revtex4-1}

\usepackage{amsmath}
\usepackage{graphicx}
\usepackage[dvipsnames]{color}
\usepackage{hyperref}
\usepackage{psfrag}
\usepackage{stmaryrd}
\usepackage{amssymb}
\usepackage{wasysym}
\usepackage{float}
\usepackage{color}

\usepackage[T1]{fontenc}
\usepackage[french,english]{babel}
\usepackage{lmodern}
\usepackage{graphicx} 
\usepackage{times}
\usepackage{amsmath,amssymb}
\usepackage{color}

\newcommand{\bnabla}{\boldsymbol{\nabla}}

\newcommand{\la}{\left<}
\newcommand{\ra}{\right>}

\begin{document}

\title{Vortex core radius in baroclinic turbulence: Implications for scaling predictions}
%


%

\author{Gabriel Hadjerci}
\email[]{gabriel.hadjerci@cea.fr}
\affiliation{Universit\'e Paris-Saclay, CNRS, CEA, Service de Physique de l'Etat Condens\'e, 91191 Gif-sur-Yvette, France.}
\author{Basile Gallet}
\affiliation{Universit\'e Paris-Saclay, CNRS, CEA, Service de Physique de l'Etat Condens\'e, 91191 Gif-sur-Yvette, France.}





\begin{abstract}
We revisit the vortex-gas scaling theory for heat transport by baroclinic turbulence based on the empirical observation that the vortex core radius departs from the Rossby deformation radius for very low bottom drag coefficient. We derive a scaling prediction for the vortex-core radius. For linear bottom drag  this scaling dependence for the vortex-core radius does not affect the vortex-gas predictions for the eddy diffusivity and mixing-length, which remain identical to those in Gallet \& Ferrari (Proc. Nat. Acad. Sci. USA, 117, 2020). By contrast, for quadratic drag the scaling dependence of the core radius induces new scaling-laws for the eddy diffusivity and mixing length when the quadratic-drag coefficient becomes asymptotically low. We validate the modified scaling predictions through numerical simulations of the two-layer model with very low quadratic-drag coefficient.
\end{abstract}


\maketitle


\section{Introduction}

The large-scale oceanic currents and atmospheric jets are in thermal wind balance with meridional buoyancy gradients. This configuration is subject to the baroclinic instability and rapidly evolves into a turbulent flow that enhances buoyancy transport in the meridional direction. The simplest model for such baroclinic turbulence is the two-layer quasi-geostrophic model (2LQG), put forward by Phillips in 1954~\cite{Phillips}. The model has been extensively described in the literature~\citep{Salmon,Salmon80,Larichev95,Held96,Arbic,Arbic2004b,Arbic2004a,Thompson06,Thompson07,Chang,Gallet2020,Gallet2021,CH21} and we only recall its main characteristics. Two immiscible layers of fluid sit on top of one another, with the lighter fluid in the upper layer. Considering (potential) temperature as the single stratifying agent for simplicity, we assume that the upper-layer fluid has a uniform temperature that is higher than the uniform temperature of the lower-layer fluid. We restrict our attention to layers that have equal depths in the rest state~\cite{Salmon}. The system is subject to rapid global rotation along the vertical direction, and fluid motion within the two shallow layers is governed by quasi-geostrophic (QG) dynamics~\cite{Pedloskybook,Salmonbook,Vallisbook}. We consider a base state where the velocity in the upper layer is $U{\bf e}_x$, while the velocity in the lower layer is $-U{\bf e}_x$. We denote with a subscript 1 (resp. 2) quantities in the upper (resp. lower) layer. The departure horizontal velocity in each layer is ${\bf u}_{1,2}(x,y,t)=-{\bnabla}\times[{\psi_{1,2}(x,y,t){\bf e}_z}]$. The flow evolution within each layer is governed by the conservation of potential vorticity $q_{1;2}$ within each layer. However, an insightful change of variables consists in introducing the barotropic streamfunction $\psi=(\psi_1+\psi_2)/2$, which represents the streamfunction of the vertically averaged flow, and the baroclinic streamfunction $\tau = (\psi_1-\psi_2)/2$ which, despite having dimension of a streamfunction, will be referred to as the `temperature' variable. Indeed, as discussed e.g. in Ref.~\cite{Gallet2021} the total baroclinic streamfunction (base state plus perturbation) $-U\,y + \tau(x,y,t)$ is a direct proxy for the vertically averaged temperature of the fluid column located at position $(x,y)$.

We include two kinds of dissipative processes: hyperdiffusion with a hyperdiffusivity $\nu$ to damp small-scale potential enstrophy in both layers, together with a bottom drag term in the lower layer to damp the kinetic energy produced by baroclinic instability. The governing equations for $\psi(x,y,t)$ and $\tau(x,y,t)$ finally read:
\begin{eqnarray}
\partial_t  (\boldsymbol{\nabla}^2 \psi) + J(\psi,\boldsymbol{\nabla}^2 \psi) + J(\tau,\boldsymbol{\nabla}^2 \tau) + U \partial_x (\boldsymbol{\nabla}^2 \tau) & = &  -\nu \boldsymbol{\nabla}^{10} \psi  + \text{drag}/2 \, ,  \label{eqpsi} \\
\partial_t  [\boldsymbol{\nabla}^2 \tau - \lambda^{-2} \tau] + J(\psi,\boldsymbol{\nabla}^2 \tau - \lambda^{-2} \tau) + J(\tau,\boldsymbol{\nabla}^2 \psi) + U \partial_x [ \boldsymbol{\nabla}^2 \psi + \lambda^{-2} \psi ] & = & -\nu \boldsymbol{\nabla}^{8} [ \boldsymbol{\nabla}^2 \tau - \lambda^{-2} \tau]  - \text{drag}/2 \, , \label{eqtau}
\end{eqnarray}
where $J(f,g)= \partial_x (f) \partial_y (g) - \partial_y (f) \partial_x (g)$, $\lambda$ denotes the Rossby deformation radius and `$\text{drag}$' denotes the drag term included in the lower-layer potential vorticity equation:
\begin{eqnarray}
\text{drag}  = \left\{ \begin{matrix}
-2 \kappa \boldsymbol{\nabla}^2 \psi_2 \ & \text{for linear drag     } \, , \\
- \mu \left[ \partial_x(|\boldsymbol{\nabla} \psi_2| \partial_x \psi_2) + \partial_y(|\boldsymbol{\nabla} \psi_2| \partial_y \psi_2)  \right] \ & \text{for quadratic drag} \, ,
\end{matrix} \right.
\end{eqnarray}
where $\kappa$ and $\mu$ denote the linear and quadratic drag coefficients, respectively.

We are interested in the solutions to equations (\ref{eqpsi}-\ref{eqtau}) inside a domain $(x,y)\in[0,L]^2$ with periodic boundary conditions in the horizontal directions, in the regime where $L$ is sufficiently large and $\nu$ is sufficiently small for the transport properties of the flow to be independent of these two parameters. Sufficiently small $\nu$ ensures that the energy dissipation is due to friction in the lower layer, the small-scale hyperdiffusive damping operator having a negligible contribution to the energy power integral. Large enough domain size $L$ ensures that the flow selects its own large scale through a balance between inverse energy transfers and frictional dissipation, the latter emergent scale being much smaller than $L$. In other words, a large enough domain prevents any condensation of the energy into a single coherent vortex dipole~\cite{Gallet2013,Frishman2018}. From a practical point of view and anticipating the results below, independence of the transport properties with respect to the domain size $L$ arises when the latter is greater than approximately six times the mixing-length estimate $\ell_2$ (the typical inter-vortex distance of the flow, see below).

We wish to characterize the transport properties of the flow as functions of the weak bottom drag coefficient $\kappa$ or $\mu$. Denoting a time and horizontal area average as $\la \cdot \ra$, the key quantity of interest is the eddy-induced diffusivity $D=\la  \tau \partial_x \psi\ra/U$, where $\la  \tau  \partial_x \psi \ra$ denotes the meridional heat flux, and $U$ is minus the meridional background temperature gradient (that is, the background gradient of the total baroclinic streamfunction, see above). 
Non-dimensionalizing time and space using the background flow velocity $U$ and the Rossby deformation radius $\lambda$, we seek the dependence of the dimensionless diffusivity $D_*=D/(U\lambda)$ on the dimensionless friction coefficient $\kappa_*=\kappa \lambda/U$ or $\mu_*=\mu \lambda$. 

Following Phillips, various authors have investigated these transport properties with or without the inclusion of a planetary vorticity gradient $\beta$. The traditional approach consists in invoking standard Kolmogorov cascade arguments~\cite{Salmon,Salmon80,Larichev95,Held96,Chang,CH21,Chen2023}. However, for low drag coefficient it was recently realized by Thompson \& Young (Ref.~\cite{Thompson06}, TY in the following) that the flow consists in a dilute gas of intense vortices that is maybe better described in physical space than in spectral space. Gallet \& Ferrari further built on this empirical observation to derive a quantitative theory for the diffusivity and mixing length, coined the vortex-gas scaling theory (Ref.~\cite{Gallet2020}, GF in the following). The vortex-gas theory leads to scaling predictions that agree better with the numerical data than cascade-like predictions~\cite{Held96}. The vortex-gas predictions have been subsequently extended to the $\beta$-plane~\cite{Gallet2021}, and they have been shown to carry over to a fully three-dimensional Eady system with linear bottom drag~\cite{Gallet2022}.

As discussed in Ref.~\cite{Chen2023} and as can be seen in GF, while the vortex-gas predictions are in excellent agreement with the numerical data for linear bottom drag, the agreement is slightly less satisfactory for quadratic bottom drag. More importantly, the agreement does not seem to improve as the quadratic bottom drag coefficient is further reduced, which questions the validity of the theory for asymptotically weak quadratic bottom drag. In this article we refine the vortex-gas theory in a way that better captures this very low quadratic drag regime.

\section{Keeping an arbitrary vortex core radius in the vortex-gas theory}

\begin{figure}
    \centerline{\includegraphics[width=6 cm]{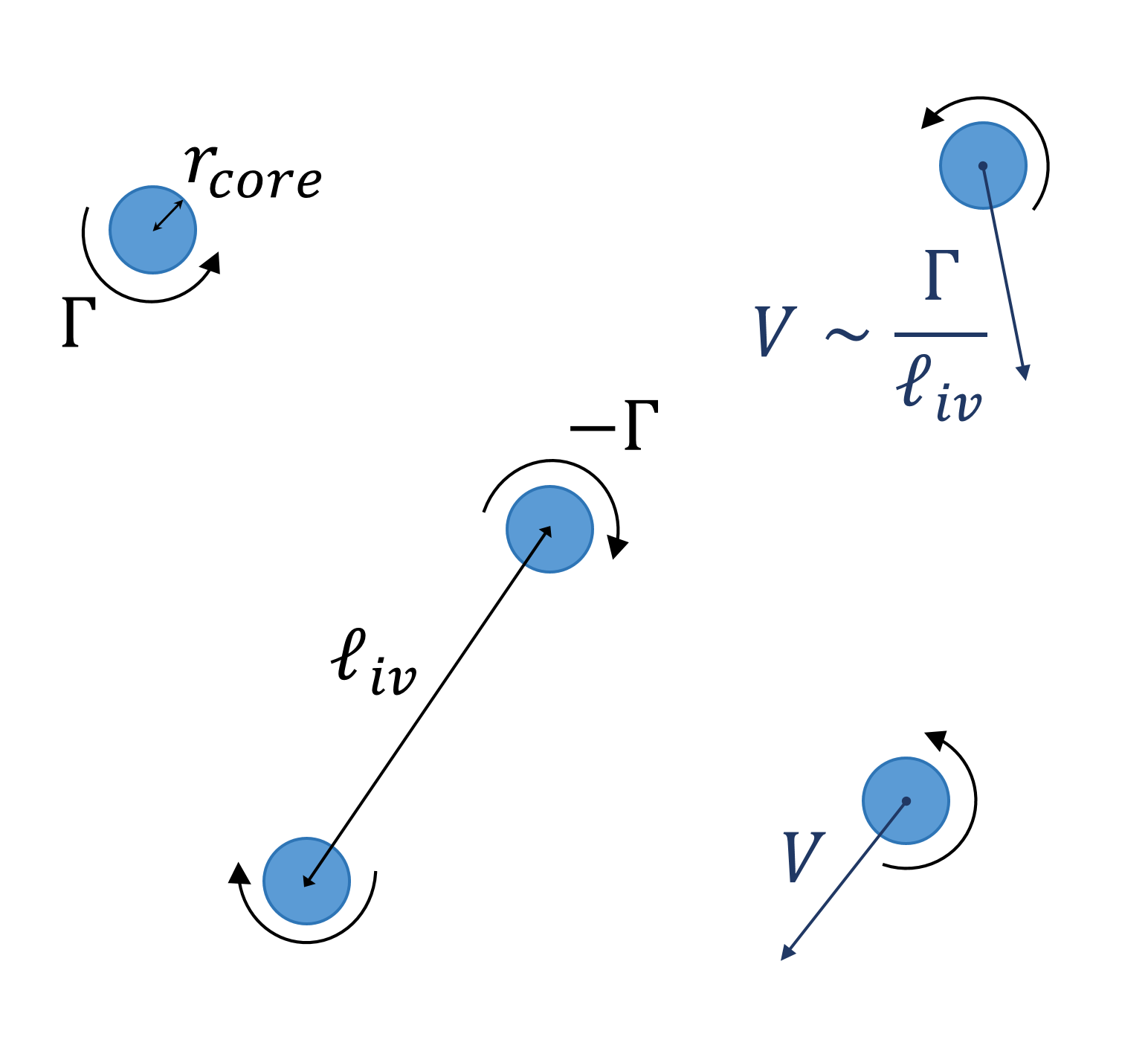} }
   \caption{In the vortex-gas theory, the barotropic flow is described as a gas of idealized vortices of circulation $\pm \Gamma$, with identical core radius $r_\text{core}$. The temperature field retains large values in the inter-vortex region, whereas the barotropic vorticity vanishes outside the vortex cores in this idealized picture. The typical magnitude of the core vorticity and temperature are denoted as $\zeta_\text{core}$ and $\tau_\text{core}$, respectively. The typical inter-vortex distance is $\ell_{iv}$, and the vortex cores wander around with a typical velocity $V \sim \Gamma/\ell_{iv}$ as a result of mutual induction. \label{fig:schematicVG}}
   \label{vortex_gas_schematic}
\end{figure}

We briefly recall the scaling arguments of the vortex-gas theory. The barotropic vorticity field is represented schematically as a dilute ensemble of vortices with circulation $\pm \Gamma$, see Figure~\ref{fig:schematicVG}. The typical inter-vortex distance is denoted as $\ell_{iv}$, and the various vortex cores move as a result of mutual induction with a typical velocity $V \sim \Gamma/\ell_{iv}$. While GF readily assume that the vortex core radius is of order $\lambda$, in the present section we retain an arbitrary vortex core radius $r_{\text{core}}$, which we write in dimensionless form as $r_*=r_{\text{core}}/\lambda$. The typical barotropic vorticity and temperature within a vortex core are denoted as $\zeta_{\text{core}}$ and $\tau_{\text{core}}$, where $\zeta=\Delta \psi$ is the barotropic vorticity.

The vortex-gas theory describes the dilute regime where $\ell_{iv} \gg r_{\text{core}}$. In line with the schematic in Figure~\ref{fig:schematicVG}, the barotropic vorticity is assumed to be nonzero inside the vortex cores only. By contrast, the theory assumes that the temperature field has significant fluctuations between the vortices. GF thus characterize transport by the vortex gas based on the strongly idealized situation of a single self-advecting vortex dipole. The dipole consists of two vortices spinning in opposite directions with circulations $\pm \Gamma$. The vortices are separated by a distance $\ell_{iv}$, and the dipole translates at constant speed $\Gamma/\ell_{iv}$ as a result of mutual induction. Together with such translating motion, the dipole induces and advects inter-vortex temperature fluctuations. Through simulations of this idealized process GF obtained the following scaling relation for the associated transport:
\begin{eqnarray}
D & \sim & \ell_{iv} V \, ,\label{D}
\end{eqnarray}
which does not involve $r_{\text{core}}$. The theory is complemented by two energetic arguments. The first energetic argument is referred to as a `slantwise' free-fall argument in GF ; consider the motion of a fluid column in the inter-vortex region, described in a similar fashion to the kinetic theory of gases. The column is initially at rest and it travels freely over a mean free path comparable to the inter-vortex distance $\ell_{iv}$, converting potential energy into kinetic energy~\cite{Green70}. Equating the initial potential energy with the final kinetic energy leads to the scaling relation ${V}/{U} \sim {\ell_{iv}}/{\lambda}$, which using (\ref{D}) can be recast as:
\begin{eqnarray}
D_* \sim (\ell_{iv}/\lambda)^2 \, . \label{freefall}
\end{eqnarray}
The second energetic argument is based on the energy power integral of the system. That is, time-averaging the energy evolution equation shows that the averaged power released by baroclinic instability equals the averaged dissipated power. Because we focus on very low hyperdiffusivity, energy dissipation is due almost entirely to bottom friction and the power integral is approximated by:
\begin{eqnarray}
 \frac{D U^2}{\lambda^2} =  \left\{ \begin{matrix}
 \kappa \la {\bf u}^2 \ra \ & \text{for linear drag     } \, , \\
 \frac{\mu}{2} \la |{\bf u}|^3 \ra \ & \text{for quadratic drag} \, .
\end{matrix} \right. \label{Ebudget}
\end{eqnarray}
The left-hand side corresponds to the rate of release of potential energy by baroclinic instability. On the right-hand side is the frictional dissipation rate of kinetic energy. Strictly speaking, because the drag force acts in the lower layer, the lower-layer velocity ${\bf u}_2$ should appear on the right-hand side of (\ref{Ebudget}).
However, because the low-drag flows are predominantly barotropic we have replaced the lower-layer velocity ${\bf u}_2$ by the barotropic velocity ${\bf u}=-{\bnabla}\times({\psi{\bf e}_z})$. The moments of the barotropic velocity field appearing in (\ref{Ebudget}) are estimated using the idealized picture of an isolated vortex located at the center of a disk of radius $\ell_{iv}$. Because the vortex is isolated, one obtains the moments of velocity field by averaging some power of the vortex velocity field over the disk of radius $\ell_{iv}$. Outside the vortex core the azimuthal velocity of the vortex is $\pm \Gamma/(2 \pi r)$, with $r$ the radius coordinate from the center of the vortex. Spatial averaging over the disk of radius $\ell_{iv}$ then leads to:
\begin{eqnarray}
\la {\bf u}^2 \ra & \sim & \frac{1}{ \ell_{iv}^2} \int_{r_\text{core}}^{\ell_{iv}} \frac{\Gamma^2}{ r} \,  {\rm d} r \sim V^2 \log \left( \frac{\ell_{iv}}{r_\text{core}} \right) \, , \label{velvariance} \\
\la |{\bf u}|^3 \ra & \sim & \frac{1}{ \ell_{iv}^2} \int_{r_\text{core}}^{\ell_{iv}} \frac{\Gamma^3}{ r^2} \,  {\rm d} r \sim V^3 \, \frac{\ell_{iv}}{r_\text{core}}  \, . \label{vel3} 
\end{eqnarray}
The scaling predictions are obtained by combining relations (\ref{D}-\ref{Ebudget}), where the right-hand side of (\ref{Ebudget}) is estimated using (\ref{velvariance}) or (\ref{vel3}). This leads to an expression for the dimensionless diffusivity $D_*$ in terms of the dimensionless core radius $r_*$ and friction coefficient $\kappa_*$ or $\mu_*$:
\begin{eqnarray}
D_*  \sim &\quad  r_*^2 \exp \left( \frac{{\cal C}}{\kappa_*} \right) & \quad\text{for linear drag} \, , \label{Dvsrcorelinear} \\
D_*  \sim & {r_*}/{\mu_*} & \quad \text{for quadratic drag} \, , \label{Dvsrcorequad}
\end{eqnarray}
where ${\cal C}$ is a constant coefficient.

\subsection{The assumption of GF: $r_* \sim 1$}

GF assume that in the strongly turbulent regime the vortex cores retain the typical scale $\lambda$ at which baroclinic instability generates flow structures in the linear regime. This assumption leads to $r_\text{core} \sim \lambda$, that is $r_* \sim 1$. Substitution into (\ref{Dvsrcorelinear}-\ref{Dvsrcorequad}) leads to the scaling predictions put forward in GF:
\begin{eqnarray}
D_*  \sim &   \exp \left( \frac{{\cal C}}{\kappa_*} \right) & \quad\text{for linear drag}  \, , \label{DvsrcorelinearGF} \\
D_*  \sim &  \frac{1}{\mu_*}\, &  \quad \text{for quadratic drag} \, . \label{DvsrcorequadGF}
\end{eqnarray}

\subsection{Immunity of the linear-drag predictions to the core-radius dependence\label{Immunity}}

The goal of the present study is to revisit the assumption $r_\text{core} \sim \lambda$. Indeed, the literature on vortex-gas dynamics indicates that the core radius can become much greater than $\lambda$ as the inverse energy transfers proceed. As a matter of fact, TY report that the vortex core radius is greater than $\lambda$. Phenomenological models based on vortex gases with merger events also lead to a vortex core radius that is greater than the scale at which energy is input, or the typical core radius of the initial condition in freely decaying situations~\cite{Carnevale,Weiss93}. The vortex-core radius increases as the inverse energy transfers proceed, and $r_*$ thus increases with $\ell_{iv}/\lambda$, or equivalently with $D_*\sim (\ell_{iv}/\lambda)^2$. In section~\ref{sec:radius} we introduce scaling arguments that lead to $r_* \sim D_*^\alpha$ with $\alpha=1/4$. Keeping the general power-law ansatz $r_* \sim D_*^\alpha$ for now, one can check that the GF scaling predictions for linear bottom drag are immune to the vortex-core-radius correction. Indeed, substitution of $r_* \sim D_*^\alpha$ into (\ref{Dvsrcorelinear}) leads to an expression identical to (\ref{DvsrcorelinearGF}) where the coefficient ${\cal C}$ is replaced by $\tilde{\cal C}={\cal C}/(1-2\alpha)$. Because ${\cal C}$ is an adjustable coefficient of the theory, so is $\tilde{\cal C}$ and the scaling predictions are exactly identical to those in GF. This robustness of the linear-drag predictions with respect to the vortex-core-radius dependence explains the excellent agreement between the numerical data and the linear-drag prediction in GF, despite the rather crude assumption $r_* \sim 1$ made by GF.

\subsection{Impact of the core-radius dependence on the quadratic-drag scaling predictions\label{Impact}}

In contrast with the linear-drag case, the quadratic-drag predictions are strongly impacted by the vortex-core-radius scaling exponent. Substitution of the power-law ansatz $r_* \sim D_*^\alpha$ into (\ref{Dvsrcorequad}) leads to the scaling prediction:
\begin{eqnarray}
D_* & \sim & \mu_*^{\frac{1}{\alpha-1}} \, . \label{Dcorr}
\end{eqnarray}
For $\alpha=0$ the prediction above reduces to the GF prediction (\ref{DvsrcorequadGF}). However, for $\alpha \neq 0$ the power-law exponents depart from the GF values and explicitly depend on $\alpha$. In the following we derive a prediction for the exponent $\alpha$ based on scaling arguments.

\begin{figure}
   \centerline{\includegraphics[width=9 cm]{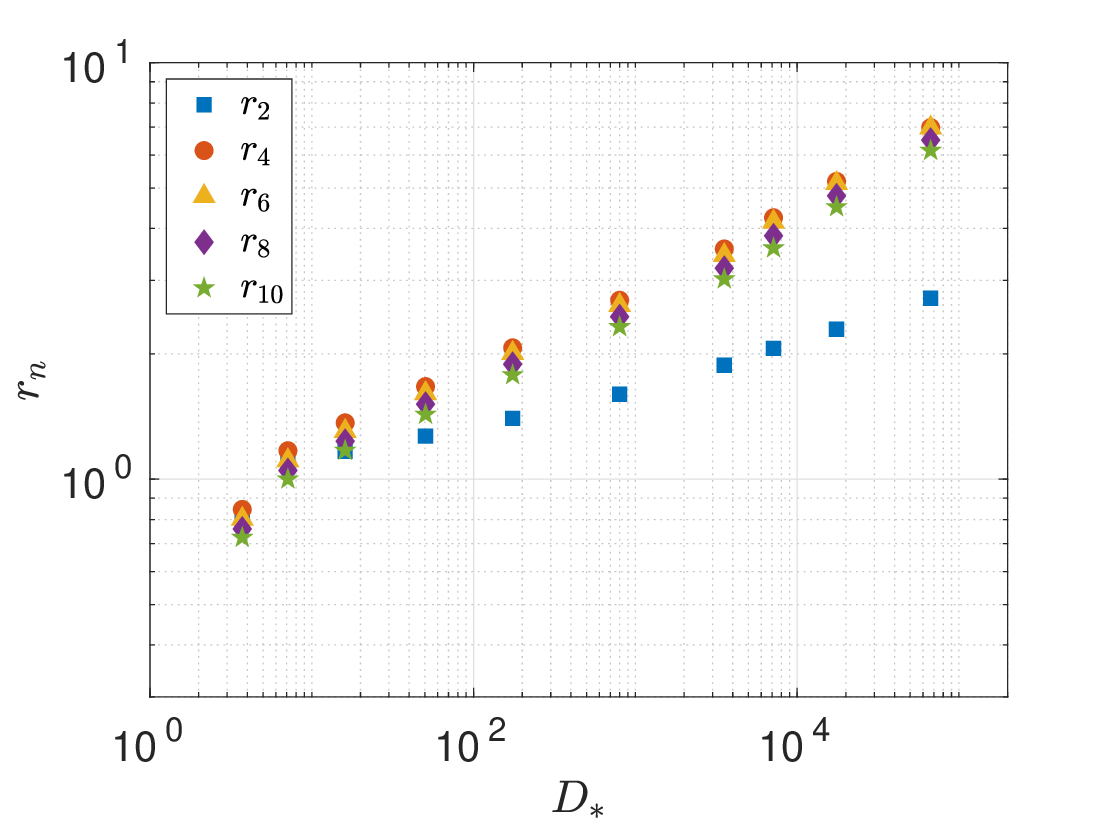} \includegraphics[width=9 cm]{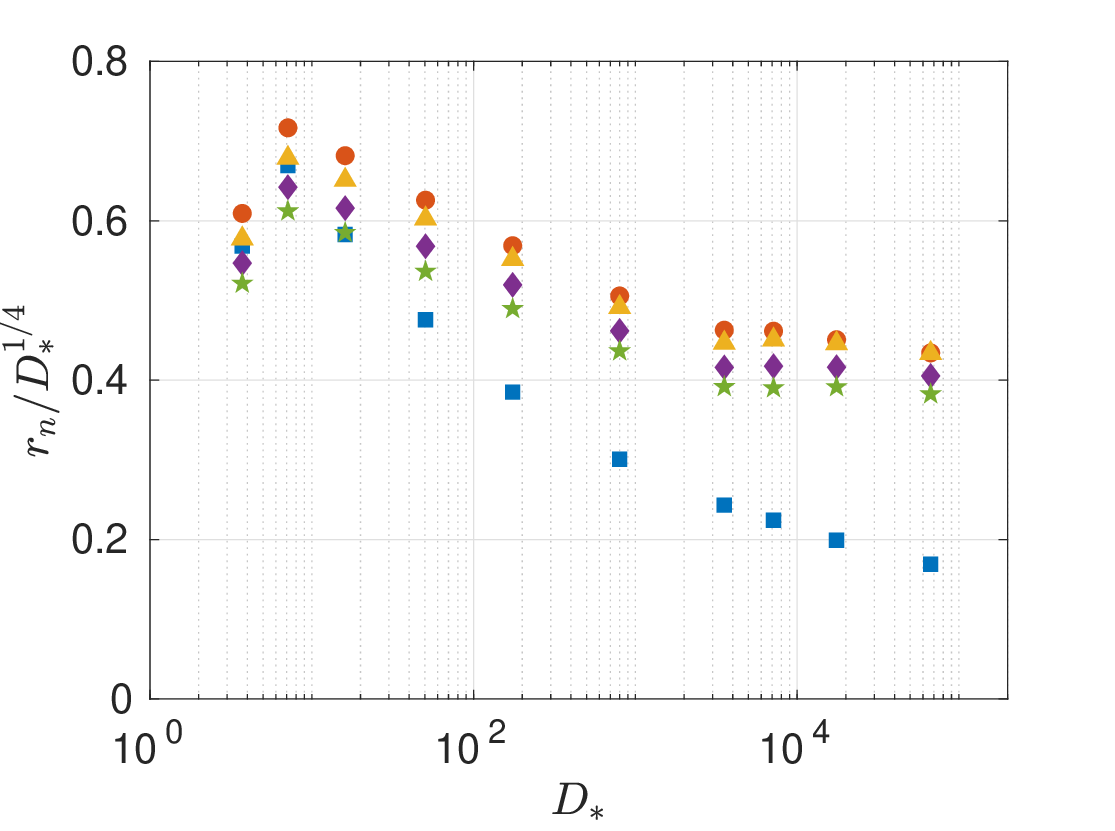} }
   \caption{Proxies $(r_n)_{n \in \{2,4,6,8,10\}}$ for the vortex-core radius, as functions of the dimensionless diffusivity $D_*$. All the proxies with $n>2$ are in excellent agreement with the scaling prediction (\ref{predictionalpha}), as illustrated by the compensated plots on the right-hand side. \label{fig:coreradius}}
\end{figure}

\subsection{A scaling prediction for the vortex core radius: $\alpha=1/4$\label{sec:radius}}

To determine the vortex-core radius, we first estimate the core vorticity $\zeta_\text{core}$. Neglecting the dissipative effects, we invoke the material conservation of the total potential vorticity within each layer, $q_1+Uy/\lambda^2$ in layer 1 and $q_2-Uy/\lambda^2$ in layer 2. With a mean free path of order $\ell_{iv}$ in the meridional direction, the fluctuations of $q_{1;2}$ are estimated as $q_{1;2} \sim U\ell_{iv}/\lambda^2$. The temperature field $\tau$ has a typical scale $\ell_{iv}$ in the inter-vortex region and $r_\text{core}$ inside the vortex cores. In the limit $\ell_{iv} \gg \lambda$ and $r_* \gg 1$ of interest here, these scales are much greater than $\lambda$. We conclude that the term $\Delta \tau$ can be neglected in the expressions of $q_{1;2}$, which reduce to:
\begin{eqnarray}
q_1 & =  & \Delta \psi_1 + \frac{\psi_2-\psi_1}{2 \lambda^2} =  \Delta (\psi+\tau) - \frac{\tau}{\lambda^2}  \simeq \zeta - \frac{\tau}{\lambda^2} \, , \label{approxq1}\\
q_2 & =  & \Delta \psi_2 + \frac{\psi_1-\psi_2}{2 \lambda^2} =  \Delta (\psi-\tau) + \frac{\tau}{\lambda^2}  \simeq \zeta + \frac{\tau}{\lambda^2} \, . \label{approxq2}
\end{eqnarray}
The total QGPV being a material invariant within each layer, $q_1$ and $q_2$ retain the same order of magnitude $U\ell_{iv}/\lambda^2$ outside and inside the vortex cores (consider for instance a fluid particle of layer $1$ that lies outside a vortex core at some initial time and inside a vortex core at some subsequent time, conserving its potential vorticity). Within a vortex core, the approximate expressions (\ref{approxq1}) for $q_1$ and (\ref{approxq2}) for $q_2$ yield respectively:
\begin{eqnarray}
\zeta_\text{core} - {\tau_\text{core}}/{\lambda^2} & \sim & U\ell_{iv}/\lambda^2 \, , \label{estimq1} \\
\zeta_\text{core} + {\tau_\text{core}}/{\lambda^2} & \sim & U\ell_{iv}/\lambda^2 \, . \label{estimq2}
\end{eqnarray}
Summing these two estimates finally leads to $\zeta_\text{core} \sim U\ell_{iv}/\lambda^2$. We recast this expression for $\zeta_\text{core}$ into an expression for the vortex core radius $r_\text{core}$ by expressing the vortex circulation as $\Gamma \sim \zeta_\text{core} r_\text{core}^2$ with $\Gamma \sim \ell_{iv} V \sim U \ell_{iv}^2 / \lambda$. Substituting the estimate for $\zeta_\text{core}$ finally leads to
\begin{eqnarray}
r_* \sim D_*^{1/4} \, , \label{predictionalpha}
\end{eqnarray}
that is $\alpha=1/4$.
As discussed in sections \ref{Immunity} and \ref{Impact} this nonzero value for $\alpha$ does not affect the scaling predictions for linear drag, but it does modify the scaling exponent of $D_*$ in the case of quadratic drag, which following (\ref{Dcorr}) becomes:
\begin{eqnarray}
D_* & \sim &  \mu_*^{-4/3} \, . \label{Dmodified}
\end{eqnarray}

\section{Numerical assessment of the modified scaling predictions}

To investigate the validity of the new scaling prediction (\ref{Dmodified}) we have performed numerical simulations of the system over an extended range of quadratic drag $\mu_* \in [10^{-4}, 1]$. The hyperviscosity coefficient is set at $\nu/(U \lambda^7)=10^{-13}$, which allows us to reach the very low drag regime while remaining close-enough to the $\nu$-independent regime (we have estimated that the values of $D_*$ reported here are typically 25\% lower than their $\nu \to 0$ limit). Additionally we make sure that the domain is large enough for $D_*$ to be close enough to its asymptotic value for an infinite domain (we have estimated that the values reported here are within 10\% of their $L \to \infty$ asymptotic limit).

\subsection{Core-radius dependence}

With the goal of testing the scaling prediction for the vortex-core radius, we first extract the various moments of the barotropic vorticity field, $\zeta_n=\la \zeta^n \ra^{1/n}$. Based on the idealized vortex-gas picture of Figure~\ref{fig:schematicVG}, where $\zeta=\zeta_\text{core}$ within the vortex cores and $\zeta=0$ outside, these moments are estimated as:
\begin{eqnarray}
\zeta_n & \sim & \left(\zeta_\text{core}^n \frac{r_\text{core}^2 }{ \ell_{iv}^2} \right)^{1/n} \sim \zeta_\text{core} {r_*^{2/n} }D_*^{-1/n} \sim \frac{U}{\lambda} r_*^{2(\frac{1}{n}-1)} D_*^{1-\frac{1}{n}} \, , \label{estimatemoments}
\end{eqnarray}
where we have used (\ref{freefall}) for the second equality and $D \sim \ell_{iv} V \sim \Gamma \sim \zeta_\text{core} r_\text{core}^2$ to express $\zeta_\text{core}$ in terms of $D_*$ and $r_*$ and obtain the last equality. From the estimate (\ref{estimatemoments}) we define a proxy $r_n$ for the dimensionless vortex-core radius $r_*$ associated with the $n$-th moment of the barotropic vorticity:
\begin{eqnarray}
r_n = \left( \frac{\lambda \zeta_n}{U} \right)^{\frac{n}{2(1-n)}} D_*^{1/2} \, .
\end{eqnarray}
To investigate the scaling behavior of the typical vortex core radius $r_\text{core}$ in the numerical simulations, we show in Figure~\ref{fig:coreradius} the proxies $r_n$ for $n$ ranging from $2$ to $10$. 
The proxies are plotted as functions of $D_*$ to investigate the validity of the scaling prediction (\ref{predictionalpha}). We observe that the proxies for $n>2$ are in excellent agreement with this scaling prediction, with $r_2$ displaying a slightly weaker scaling exponent. This may be an indication that the scaling theory provides a good description of the strongest vortices that populate the barotropic flow. In other words, the idealized identical vortices in Figure~\ref{vortex_gas_schematic} should be thought of as the strongest vortices of the barotropic flow. The sea of weaker vortices that coexists with these strong isolated vortices only affects the low-order moment $\zeta_2$, and thus $r_2$. By contrast, higher-order moments are predominantly sensitive to the strong vortices and display good agreement with the vortex-gas scaling prediction.

\subsection{Diffusivity}

\begin{figure}
    \centerline{\includegraphics[width=9 cm]{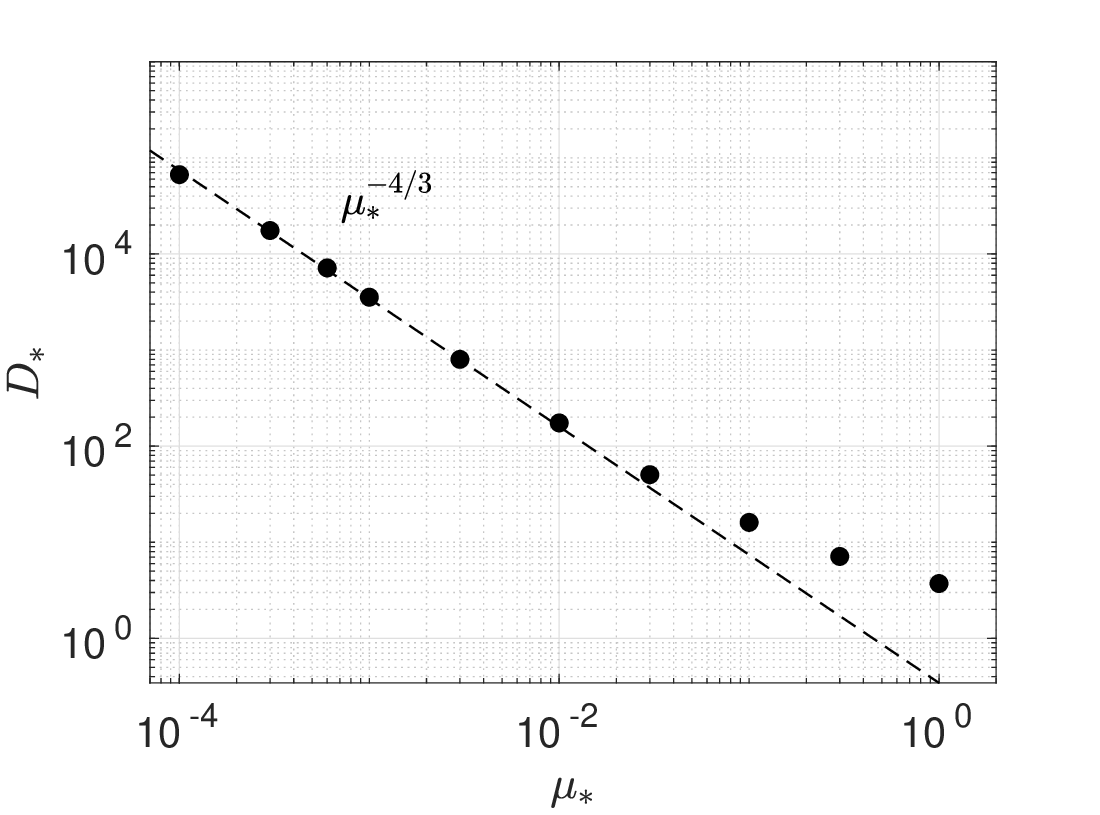} \includegraphics[width=9 cm]{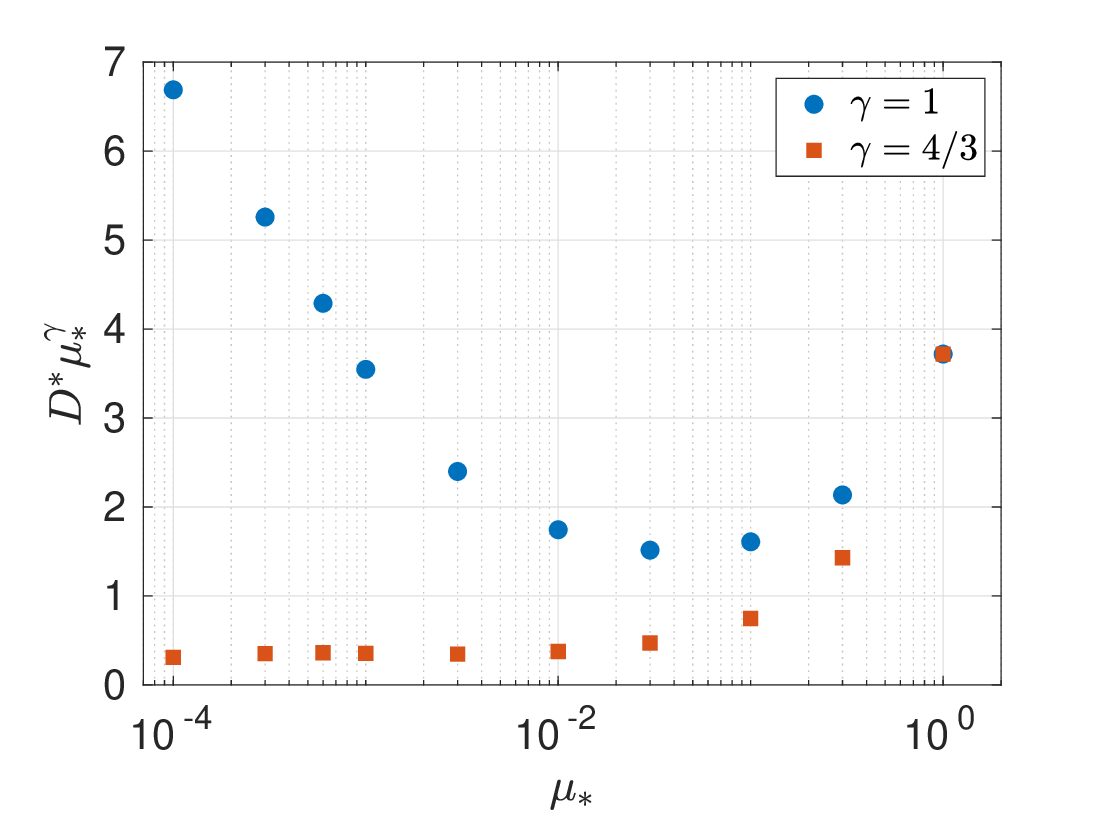} }
   \caption{Diffusivity $D_*$ as a function of the quadratic drag coefficient $\mu_*$. The data points are in reasonable agreement with the GF prediction (\ref{DvsrcorequadGF}) for moderately low drag. For asymptotically low drag they switch to the scaling behavior (\ref{Dmodified}), as illustrated by the compensated plots on the right-hand side. \label{fig:Dvsmu}}
\end{figure}

Having validated the scaling exponent $\alpha=1/4$ for the vortex-core radius dependence, we plot in Figure~\ref{fig:Dvsmu} the eddy-induced diffusivity $D_*$ as a function of the quadratic drag coefficient $\mu_*$ in the very-low drag regime. We also provide plots of $D_*$ compensated by the GF prediction (\ref{DvsrcorequadGF}) and by the new prediction (\ref{Dmodified}) obtained by including the scaling dependence of the vortex-core radius. While the data agree reasonably with the GF prediction for moderately low drag, say $\mu_* \in [3 \times 10^{-3}, 0.3]$, they depart from it for an even lower drag coefficient. In the very-low drag asymptotic regime, which begins around $\mu_* \lesssim 3 \times 10^{-3}$, the numerical values of $D_*$ are in excellent agreement with the modified scaling prediction (\ref{Dmodified}), with a best-fit exponent $D_* \sim \mu_*^{-1.303}$. One can check that this range of $\mu_*$ is also where the vortex-core-radius scaling prediction is accurately satisfied, see Figure~\ref{fig:coreradius}.

\subsection{Inter-vortex distance and mixing length}

\begin{figure}
    \centerline{\includegraphics[width=9 cm]{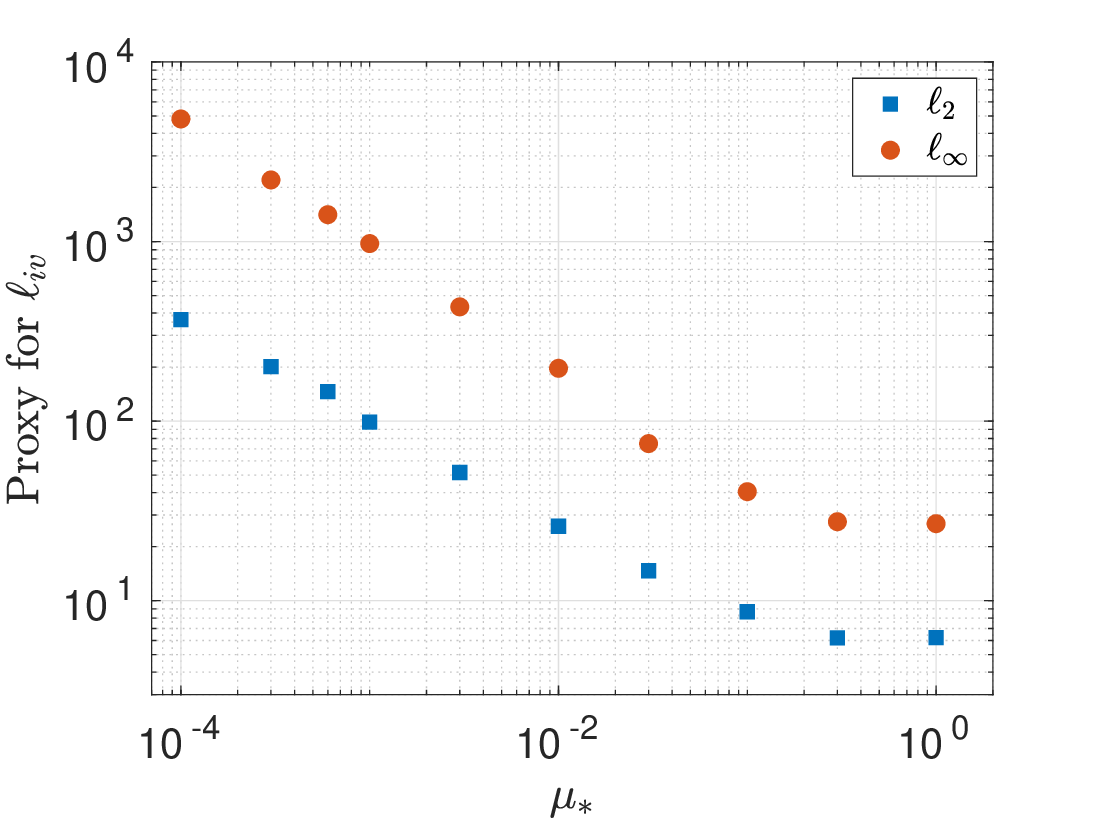} \includegraphics[width=9 cm]{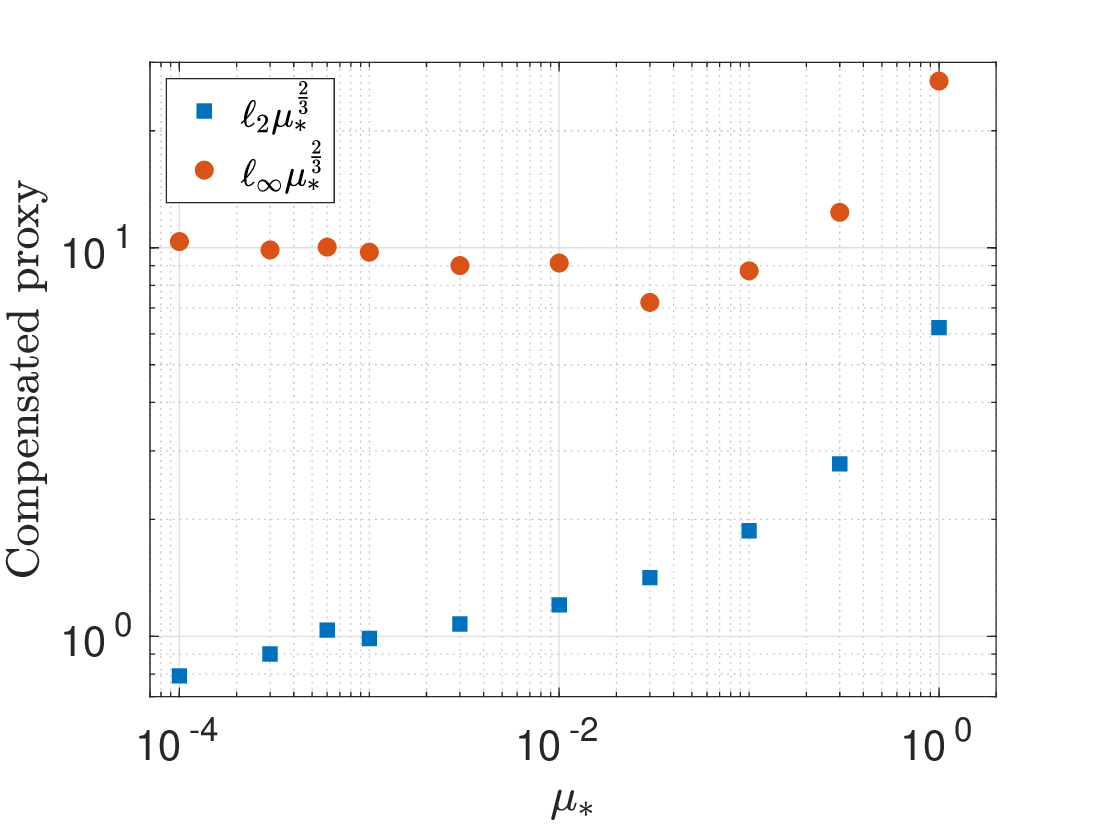} }
   \caption{The two proxies $\ell_2$ and $\ell_\infty$ for the inter-vortex distance as functions of the quadratic drag coefficient $\mu_*$. The agreement with the scaling prediction (\ref{ellmodified}) is very good in the asymptotically low drag regime, as illustrated by the compensated plots on the right-hand side. \label{fig:ellvsmu}}
\end{figure}

The present theory also provides a modified scaling prediction for the dependence of the inter-vortex distance on quadratic drag. Combining (\ref{freefall}) with (\ref{Dmodified}) leads to:
\begin{eqnarray}
\frac{\ell_{iv}}{\lambda} \sim \mu_*^{-2/3} \, . \label{ellmodified}
\end{eqnarray}
As for the core radius, various proxies can be defined for the inter-vortex distance, this time based on the various moments $\tau_n=\la |\tau|^n \ra^{1/n}$ of the temperature field $\tau$.
Following TY, GF used a dimensionless mixing-length $\ell_2 = \tau_2/(U \lambda)$ as a proxy for the inter-vortex distance ($\ell_2$ is denoted as $\ell_*$ in GF). This is arguably the simplest characteristic length associated with the fluctuating temperature field. In Figure~\ref{fig:ellvsmu} we show that the scaling dependence of $\ell_2$ with $\mu_*$ is indeed reasonably well captured by the scaling prediction (\ref{ellmodified}), which in the very-low-drag regime constitutes an improvement as compared to the GF prediction ${\ell_{iv}}/{\lambda} \sim \mu_*^{-1/2}$. That being said, the scaling exponent of $\ell_2$ with $\mu_*$ is slightly shallower than $-2/3$ over the last decade in $\mu_*$, with a best-fit exponent of the order of $-0.56$. The agreement remains very satisfactory and may improve as one reaches even lower values of the drag coefficient. However, as discussed above the idealized vortex-gas picture seems to describe predominantly the strongest vortices within the barotropic flow. This leads one to define the alternate proxy $\ell_\infty=\tau_\infty/(U \lambda)$, which may be more directly related to the typical inter-vortex distance $\ell_{iv}$ than $\ell_2$. Indeed, $\tau_\infty=\lim_{n\to \infty} \tau_n$ is the infinite norm (the time average of the maximum over $x$ and $y$ of the absolute value) of the temperature field $\tau$. It thus senses the core temperature of the strongest vortex inside the domain, providing a proxy for the typical core vorticity $\tau_\text{core}$ of the vortex gas. Subtracting the two estimates (\ref{estimq1}) and (\ref{estimq2}) then readily leads to $\tau_\text{core}/\lambda^2 \sim U \ell_{iv}/\lambda^2$, which using $\tau_\text{core} \sim \tau_\infty$ can be recast as:
\begin{eqnarray}
\ell_\infty \sim \frac{\ell_{iv}}{\lambda} \, .
\end{eqnarray}
That is, we expect the proxy $\ell_\infty$ to faithfully obey the scaling prediction for the dimensionless inter-vortex distance ${\ell_{iv}}/{\lambda}$. In Figure~\ref{fig:ellvsmu}, we validate this prediction by plotting $\ell_\infty$ as a function of $\mu_*$. The data points show very good agreement with the prediction (\ref{ellmodified}), with a best-fit exponent $-0.69$ over the last decade in $\mu_*$.

\section{Conclusion}

We have derived a scaling prediction for the typical radius of the vortex cores arising in equilibrated low-drag baroclinic turbulence. For linear bottom drag the vortex-gas scaling prediction for the eddy-induced diffusivity is immune to this core-radius dependence and remains identical to the prediction in GF. By contrast, for quadratic bottom drag the scaling predictions are modified when including the dependence of the vortex-core radius. We have validated the new scaling predictions through numerical simulations of the 2LQG model with very low quadratic drag.

From a physical point of view it is very satisfactory that the theory captures the very-low-drag strongly turbulent asymptotic regime for both linear and quadratic bottom drag, with possible relevance to exoplanetary oceans and atmospheres. In the context of the parameterization of mesoscale turbulence in the Earth's ocean, dissipation on the ocean floor is sometimes modeled as a linear friction force, but more often as quadratic drag. The drag coefficient is believed to be only moderately small, however, and mesoscale ocean turbulence appears as a moderately dilute vortex gas~\cite{Arbic2004a,Arbic2004b,Venaille2011} for which the predictions in GF are likely sufficient.

\acknowledgments This research is supported by the European Research Council under grant agreement FLAVE 757239. We thank W.R. Young for insightful comments. 

\end{document}